\documentclass[prb,aps,twocolumn,showpacs,floats]{revtex4}

\usepackage{epsfig,bm,epsf,float,graphics,graphicx}
\usepackage{pstricks} 

\begin{document}
\draft
\title{SPIN TRANSPORT IN MAGNETIC MULTILAYERS}
\author{K. Akabli, H. T. Diep\footnote{ Corresponding author, E-mail:
diep@u-cergy.fr} and S. Reynal\footnote{Permanent address:
ENSEA, 6, Avenue du Ponceau, 95014 Cergy-Pontoise Cedex, France.}}
\address{Laboratoire de Physique Th\'eorique et Mod\'elisation,
CNRS-Universit\'e de Cergy-Pontoise, UMR 8089\\
2, Avenue Adolphe Chauvin, 95302 Cergy-Pontoise Cedex, France}

\begin{abstract}
We study by extensive Monte Carlo simulations the transport of
itinerant spins travelling inside a multilayer composed of three
ferromagnetic films antiferromagnetically coupled to each other in
a sandwich structure.  The two exterior films interact with the
middle one through non magnetic spacers.  The spin model is the
Ising one and the in-plane transport is considered. Various
interactions are taken into account.  We show that the current of
the itinerant spins going through this system depends strongly on
the magnetic ordering of the multilayer:  at temperatures $T$
below (above) the transition temperature $T_c$, a strong (weak)
current is observed. This results in a strong jump of the resistance across
$T_c$. Moreover, we observe an anomalous variation, namely a peak,
of the spin current
 in the critical region just above $T_c$. We show that  this peak is due to the formation of
domains in the temperature region between the low-$T$ ordered
phase and the true paramagnetic disordered phase.  The existence
of such domains is known in the theory of critical phenomena. The
behavior of the resistance obtained here is compared to a recent
experiment. An excellent agreement with our physical
interpretation is observed. We also show and discuss effects of
various physical parameters entering our model such as interaction
range, strength of electric and magnetic fields and magnetic film
and non magnetic spacer thicknesses.

\end{abstract}
\maketitle
\section{Introduction}

The so-called giant magnetoresistance (GMR) was discovered
experimentally twenty years ago by Fert and
coworkers\cite{Baibich} and by Grunberg\cite{Grunberg}. Since
then, intensive investigations, both experimentally and
theoretically, have been carried out to understand the origin and
the behaviors of the spin current in magnetic multilayer
systems.\cite{Fert,review,Gravier,Stewart,Monchesky,McKenna,Carva}
This spectacular development is due mainly to an important number
of industrial applications using such systems in data storage and
magnetic sensors.\cite{Fert,review}

Experimental observations show that when the spin of an itinerant
spin is parallel to the spins of the environment it will go
through easily while it will be stopped if it encounters an
antiparallel spin medium.  The resistance is stronger in the
latter case resulting in a GMR.  Although many theoretical
investigations have been carried out, detailed understanding of
the influence of each physical parameter on the spin current is still lacking. For
example the role of interface scattering and the effect of interface
roughness on the GMR are still currently investigated (see for
example Refs.\onlinecite{Monchesky,Stewart} and references therein). In
addition, to date no Monte Carlo (MC) simulations have been
performed regarding the temperature dependence of the dynamics of spins participating in the current.
This defines the aim of this work.

This paper deals with the transport of spins in a system composed
of three magnetic films.  We show that the spin current depends on
the orientation of the lattice spins found on the trajectory. The
dependence of the spin transport on the magnetic ordering, i.e., on
the temperature is studied.  The difficulty of the task is that we
have to deal at the same time with surface and interface effects and with
dynamical properties of itinerant spins interacting with the
lattice spins. Surface physics of systems such as films and
multilayers have been enormously studied at equilibrium during
the last 30 years. This was motivated in particular by applications in
magnetic recording, let alone fundamental theoretical interests.
Much is understood theoretically and experimentally in  thin films
whose surfaces are 'clean', i.e., contain no impurities, no steps etc.
\cite{zangwill,bland-heinrich,Binder-surf,Diehl,ngo2004,ngo2007} Far less is
known --- at least theoretically --- on complicated thin films with
special surface conditions such as defects, arrays of dots and
magnetization reversal phenomena.  As a result, studying the behavior of
itinerant electrons injected into such systems is a formidable task
which cannot be fulfilled in every respect.

The paper is organized as follows. Section II is devoted to the
description of our model and the rules that govern its dynamics.
We take into account (i) interactions between itinerant and
lattice spins, and (ii) interactions between itinerant spins
themselves (iii) interactions between lattice spins. Where rules
governing the dynamics are concerned, we include a thermodynamic
force due to the gradient of itinerant spin concentration, an
applied electric field that drives electrons, and the effect of a
magnetic field. In section III, we describe our MC method and
discuss the results we obtained for several physical quantities in
various situations, e.g., the mean free-path, the spin current and
the resistance. Comparison with a very recent
experiment\cite{Brucas} performed on a permalloy-insulator
multilayer is also shown in this section. Concluding remarks are
given in Section IV.

\section{Model}
\subsection{Interactions}

\begin{figure}[htb!]
\includegraphics[width=8.0cm]{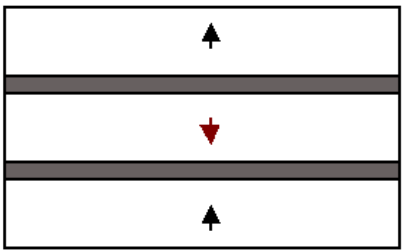}
\caption{Ground state spin configuration.  Thick arrows indicate
the spin orientations.  Nonmagnetic spacers are black. }
\label{fig:gsstruct}
\end{figure}

We consider in this paper three ferromagnetic films being
antiferromagnetically coupled to each other via nonmagnetic
layers. We use the Ising model and the face-centered cubic (FCC)
lattice for the films.  The system is shown in Fig.
\ref{fig:gsstruct} where the films are stacked along the $z$
direction.

The multilayer is made up of three films each of which has a volume given by
 $N_x\times N_y \times N_z$, where $N_z$ denotes the number of
atomic layers (i.e., single film thickness).  Periodic boundary
conditions are used in the $xy$ planes. Nonmagnetic spacers
sandwiched between films have a thickness $d$.

\begin{figure}[htb!]
\input{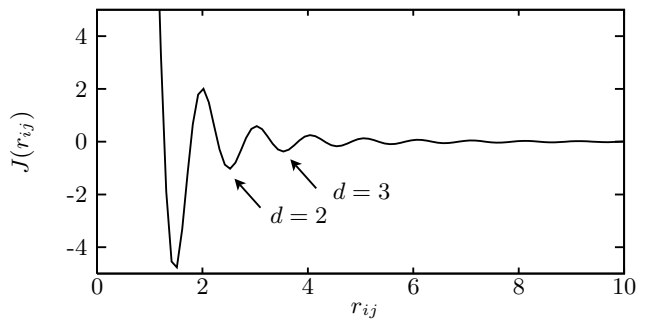} \caption{RKKY interaction for two spins across
the nonmagnetic layer is shown as a function of their distance
$r_{ij}$. $J_0=16.6$ and $\alpha =6.18$ have been used so that
$J_{i,j}$=-1.005 for nonmagnetic spacer thickness $d=2$.}
\label{fig:RKKY}
\end{figure}

 Spins at FCC lattice sites are called "lattice spins" hereafter.  They interact with each other
through the following Hamiltonian:

\begin{equation}
\mathcal H_l=-\sum_{\left<i,j\right>}J_{i,j}\mathbf S_i\cdot\mathbf S_j,  \label{eqn:hamil1}
\end{equation}
where $\mathbf S_i$ is the Ising spin at lattice site $i$,
$\sum_{\left<i,j\right>}$ indicates the sum over every nearest-neighbor (NN) spin pair
$(\mathbf S_i, \mathbf S_j)$. For simplicity, we will consider
the case where all exchange interactions $J_{i,j}$ are
ferromagnetic and equal to $J (>0)$, except for the interaction across
the nonmagnetic spacer which we define using the following
RKKY model,
\begin{equation}\label{RKKY}
J_{i,j}= J_0\frac{\cos (\alpha r_{ij})}{r_{ij}^3}.
\end{equation}
Here, $i$ and $j$ refer to spins on either side of a nonmagnetic layer,
and $J_0$ and $\alpha$ are constants chosen in such a way that the strength
of $J_{i,j}$ is physically reasonable. The shape of the interaction is sketched in Fig.
\ref{fig:RKKY}.

When the coupling across nonmagnetic layers is
antiferromagnetic, the ground state corresponds to the two exterior films having
spins pointing in one direction and the interior one spins pointing in the opposite direction.

In order to study the spin transport inside the multilayer system
described above, we consider a flow of itinerant spins interacting
with each other and with the lattice spins.  The interaction
between itinerant spins is defined  as follows,

\begin{equation}
\mathcal H_m=-\sum_{\left<i,j\right>}K_{i,j}\mathbf
s_i\cdot\mathbf s_j,  \label{eqn:hamil2}
\end{equation}
where $\mathbf s_i$ is the Ising spin at position  $\vec r_i$,
and $\sum_{\left<i,j\right>}$ denotes a sum over every spin pair
$(\mathbf s_i, \mathbf s_j)$.  The interaction $K_{i,j}$
depends on the distance between the two spins, i.e.,
$r_{ij}=|\vec r_i-\vec r_j|$.  A specific form of $K_{i,j}$ will
be chosen below.  The interaction between itinerant spins and
lattice spins is given by

\begin{equation}
\mathcal H_r=-\sum_{\left<i,j\right>}I_{i,j}\mathbf
s_i\cdot\mathbf S_j,  \label{eqn:hamil3}
\end{equation}
where the interaction $I_{i,j}$ depends on the distance
between the itinerant spin $\mathbf s_i$ and the lattice spin
$\mathbf S_i$. For the sake of simplicity, we assume the same form
for $K_{i,j}$ and $I_{i,j}$, namely,
\begin{eqnarray}
K_{i,j}&=& K_0\exp(-r_{ij})\\
I_{i,j}&=& I_0\exp(-r_{ij}),
\end{eqnarray}
where $K_0$ and $I_0$ are constants expressing the respective
strength of interactions.

\subsection{Dynamics}

\begin{figure}[htb!]
\input{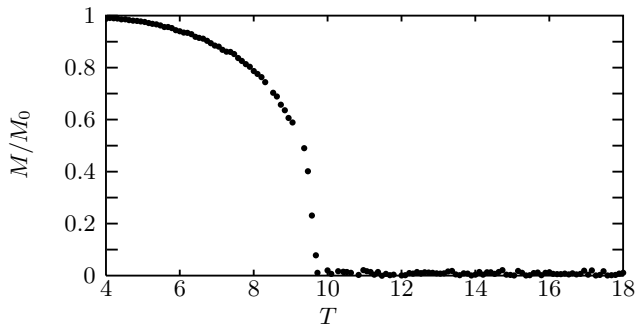} \caption{Total staggered lattice magnetization
versus temperature $T$. $M_0$ is the ground state staggered
lattice magnetization. $T_c$ is $\simeq 9.75$ in unit of $J=1$.}
\label{fig:M(T)}
\end{figure}

Let us now explain the procedure we utilize in our simulation.
First we study the thermodynamic properties of the multilayer
system alone, i.e., without itinerant spins, using Eq. (\ref
{eqn:hamil1}).  In this view, we perform MC simulations in order
to determine quantities as the internal energy, the specific heat,
layer magnetizations, the susceptibility, ... as  functions of
temperature $T$.\cite{Binder} From these physical quantities we
determine the critical temperature $T_c$ below which the system is
in the ordered phase, e.g., with up-spin phase for the outer films
and down-spin phase for the middle film. The total staggered
lattice magnetization is defined as $M=(M_1-M_2+M_3)/3$ where
$M_i(i=1,2,3)$ is the magnetization of the $i$th film. We depict
in Fig.\ref{fig:M(T)} the lattice magnetization versus $T$.

\begin{figure}[htb!]
\input{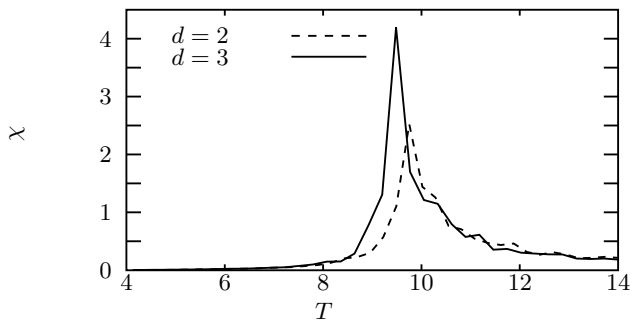}
\caption{Susceptibility $\chi$ of staggered lattice magnetization
versus temperature $T$ for two spacer thicknesses $d=2$ (solid line)
and $d=3$ (dashed line).} \label{fig:chi}
\end{figure}

Figure~\ref{fig:chi} shows the susceptibility calculated from the
fluctuations of $M$ for two spacer thicknesses $d=2,3$. $T_c$ is
equal to $\simeq 9.75$ and $\simeq 9.49$ respectively for these
spacers.

Once the lattice has been equilibrated at $T$, we study the dynamics of
itinerant spins at that temperature by injecting itinerant spins
with a density $n$ into the multilayer system. There are two ways
of doing this: i) the itinerant spins move parallel to the film
surface (CIP case); ii) the itinerant spins move perpendicular to
the films (CPP case). In this paper we show results in the CIP
case.

The itinerant spins in the CIP case move into the system at one end,
travel in the $x$ direction, escape the system at the other end
to reenter again at the first end under periodic boundary
conditions (PBC).  Note that PBC are used  to ensure that the
average density of itinerant spins remains constant during the
time (stationary regime). The dynamics of itinerant spins is
governed by the following interactions:

i) an electric field $\mathbf E$ is applied in the $x$ direction.
Its energy is given by
\begin{equation}
\mathcal {H}_E=-\mathbf E \cdot \mathbf v,
\end{equation}\
where $ \mathbf v$ is the velocity of the itinerant spin

ii) a chemical potential term which depends on the concentration
of itinerant spins within a sphere of radius $D_2$ ("concentration
gradient" effect). Its form is given by
\begin{equation}
\mathcal {H}_{c}= Dn(\mathbf r),
\end{equation}
 where $n(\mathbf r)$ is the concentration of
itinerant spins in a sphere of radius $D_2$ centered at $\mathbf
r$. $D$ is a constant taken equal to $K_0$ for simplicity.

iii) interactions between a given itinerant spin and
lattice spins inside a sphere of radius $D_1$
(Eq.~\ref{eqn:hamil3}).

iv) interactions between a given itinerant spin and other
itinerant spins inside a sphere of radius $D_2$
(Eq.~\ref{eqn:hamil2}).

Let us first consider the case without an applied magnetic field.

The simulation is carried out as follows: at a given $T$ we
calculate the energy of an itinerant spin by taking into account all the
interactions described above.  Then we tentatively move the spin under
consideration to a new position with a step of length $v_0$ in an
arbitrary direction. Note that this move is immediately rejected
if the new position is inside a sphere of radius $r_0$
centered at a lattice spin or an itinerant spin. This excluded space
emulates the Pauli exclusion principle in the one hand,
and the interaction with lattice phonons on the other hand.
For the sake of example, if the spacing between NN lattice spins is $\sqrt{2}$
then $r_0$ is taken of the order of 0.05. This value can be made
temperature-dependent to account for the cross section of phonon-electrons
collisions.

If the new position does not lie in a forbidden region of space, then
the move is accepted with a probability given by the standard Metropolis algorithm \cite{Binder} ;
in particular, it is always accepted if the energy of the itinerant spin at the new
position is lower than its previous value.

\section{Monte Carlo results}

In this paragraph, we show the results obtained by MC simulations
with the Hamiltonians given above. All Ising spins are of
magnitude $s=S=1$.

The parameters we use in most calculations are, except otherwise stated,
$N_x=36$, $N_y=10$ and $N_z=5$ for the dimension of the films,
$d=2$ for the spacer thickness. We also make use of PBC in the $xy$ plane.

At each temperature the equilibration time for the lattice spins
lies around $10^6$ MC steps per spin and we compute statistical
averages over $10^6$ MC steps per spin. Taking $J=1$, we obtain
$T_c\simeq 9.75$ for the estimate of the critical temperature of
the lattice spins (see Figs.\ref{fig:M(T)} and \ref{fig:chi}).
Before calculating the mean free path and the spin current, we let
$n$ itinerant spins travel through the system several thousands
times until a steady state is reached. The parameters used for the
results shown below are $D_1=D_2=1$ (in unit of the FCC cell
length), $K_0=I_0=2$, $n=1500$, $v_0=1$, $r_0=0.05$.

\begin{figure}
\input{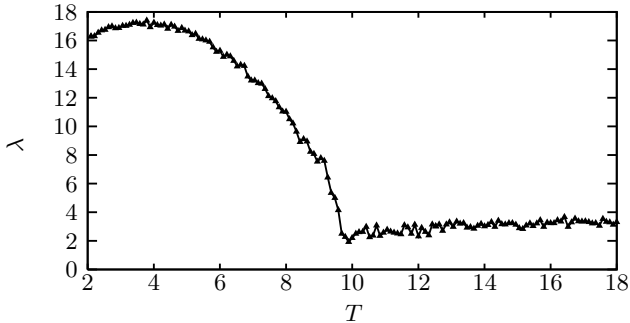}
\caption{Mean free path  $\lambda$ in unit of the FCC cell length
versus temperature $T$ , for 100 MC steps.}\label{fig:MFP}
\end{figure}

In Fig. \ref{fig:MFP} we sketch the travelling length $\lambda$
computed after a fixed lapse of time as a function of temperature
$T$. As can be seen, $\lambda$ is very large at $T<T_c$. We note
that there is a small depression in the transition region. We will
show below that this has important consequences on the spin
current.  We also note that at very low $T$ ($T< 4$), the mean free
path suffers a decrease with decreasing $T$. This is a well-known
artefact of MC simulation at very low $T$: the moving probability
is so small that the motion of itinerant spins is somewhat slowed
down. As we will show below  when comparing with experimental
data,  this freezing is also observed in real systems due to
finite experimental observation time.

\begin{figure}
\input{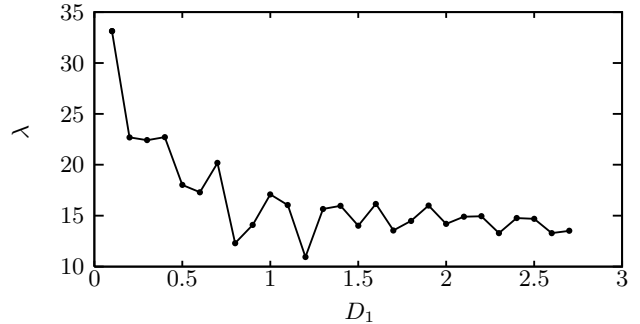}
\caption{Mean free path $\lambda$ versus  $D_1$, at $T=1$, $D_2=1$
and $E=1$.}\label{fig:D1}
\end{figure}

\begin{figure}
\input{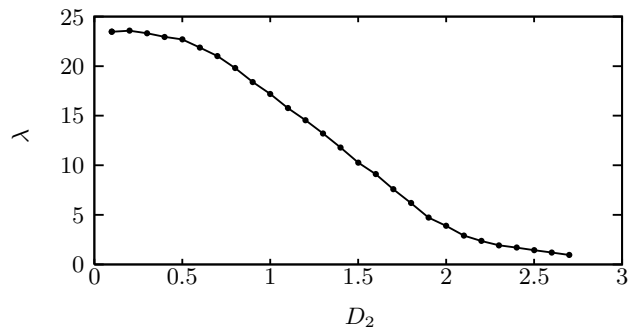}
\caption{Mean free path  $\lambda$ versus $D_2$, at $T=1$, $D_1=1$
and $E=1$.}\label{fig:D2}
\end{figure}

Figures \ref{fig:D1} and \ref{fig:D2} show the effects of varying
$D_1$ and $D_2$ at a low temperature $T=1$. As seen in Fig.
\ref{fig:D1}, $\lambda$ is
 very large at small $D_1$: this can be explained by the fact that for such small
 $D_1$, itinerant spins do not "see" lattice spins in their interaction sphere so they move almost
 in an empty space.  The effect of $D_2$ is on the other hand qualitatively very different from
 that of $D_1$ as
 seen in Fig.  \ref{fig:D2}: $\lambda$ is
 saturated at small $D_2$ and decreases to the minimum value, namely $\lambda$=1,  at
 large $D_2$.  We conclude that both $D_1$ and $D_2$ dominate $\lambda$
 at their small  values. However, at large values, only $D_2$ has
 a strong effect on $\lambda$. This effect comes naturally from the criterion on the itinerant spins concentration
 used in the moving procedure.

\begin{figure}
\input{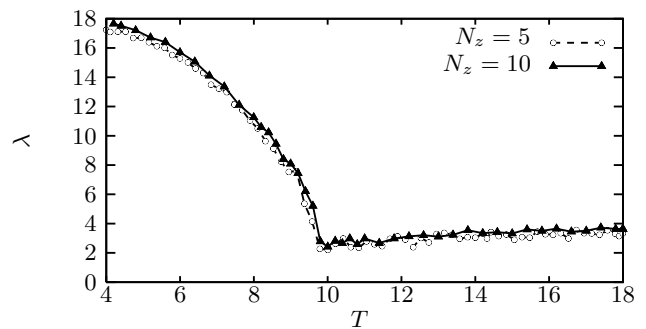} \caption{Mean free path  $\lambda$ versus $T$,
for several thickness values of the magnetic film, with
$D_1=D_2=1$ and $E=1$.}\label{fig:Nz}
\end{figure}

The mean free path is shown in Fig. \ref{fig:Nz} as a function of
$T$ for two magnetic film thicknesses.  In the absence of
interface impurities, it is expected that there would be no large
effects on the motion of itinerant spins. This is indeed what we
observe here.  Note however that the mean free path for the
smaller magnetic film thickness is systematically smaller than
that of the thicker film.  We will discuss on the role of
interfaces below while showing the resistance (Fig. \ref{resist}).

\begin{figure}
\input{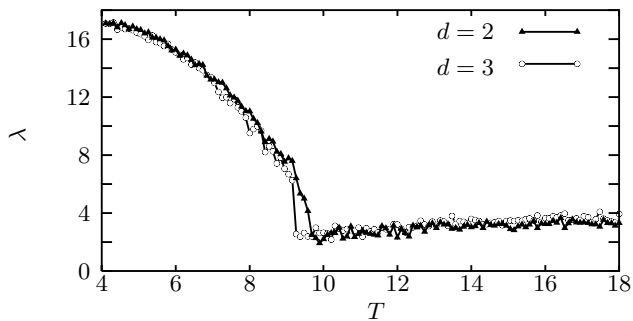}
\caption{Mean free path  $\lambda$ versus $T$, for several spacer
thicknesses with $D_1=D_2=1$ and $E=1$.}\label{fig:Nm}
\end{figure}

We show in Fig. \ref{fig:Nm} the effect of the spacer thickness on
the mean free path. Note that for each thickness value we have
used the inter film coupling constant $J_i$ calculated by Eq.
\ref{RKKY}. Increasing the thickness, i. e. decreasing $J_i$, will result in a decrease
of the mean free path visible at low $T$ as can be seen in Fig.
\ref{fig:Nm}. This is expected since the itinerant spins at
magnetic-nonmagnetic interfaces have weaker inter film coupling
energy, so they are scattered more easily.

We show in Fig. \ref{fig:E} the effect of the electric field $E$
for  $T$ both above and below $T_c$. The low-field part verifies the
Ohm regime.

\begin{figure}
\input{Figure10.pst}
\caption{Mean free path $\lambda$ versus $E$, below and above
$T_c$,  with $D_1=D_2=1$.}\label{fig:E}
\end{figure}

For the $i$-th layer, we define the resistivity as
\begin{equation}
\rho_i=\frac{1}{n_i},
\end{equation}
where $n_i$ is the number of spins crossing a unit area
perpendicular to the $x$ direction per unit of time. Note that
this definition is applied to three magnetic ($i=1,3,5$) and two
nonmagnetic layers ($i=2,4$).  The total resistance $R$ is defined
as

\begin{equation}
R^{-1}=\sum_{i=1}^5 \frac{1}{\rho_i}.
\end{equation}
This definition is suitable for low-$T$ phase where the spin
current is distinct in magnitude between magnetic and nonmagnetic
layers.  On the contrary, in the paramagnetic phase the spin
current is almost spatially uniform, and the resistance can be
defined as
\begin{equation}
R^{-1}=\frac{1}{\rho}=\frac{1}{5} \sum_{i=1}^5 n_i.
\end{equation}

In Fig.~\ref{resist} we show the resistance $R$ as a function
of temperature.
\begin{figure}
\input{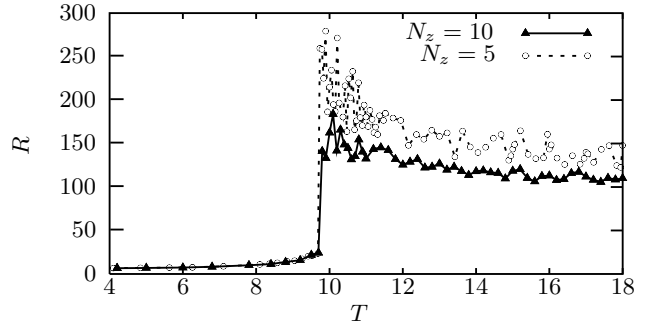} \caption{Resistance $R$ in arbitrary unit versus
temperature $T$ for two magnetic layer thicknesses.}\label{resist}
\end{figure}

There are several striking points:

\begin{itemize}
\item  $R$ is very low in the ordered phase and large in the
paramagnetic phase. Below the transition temperature, there exists
a single large cluster with small-sized excitation inside it (see
Fig. \ref{fig:meansize}), so that any itinerant spin having the
appropriate orientation goes through the structure without any
hindrance. The resistance is thus very small.

\item $R$ exhibits a cusp at the transition temperature, the
existence of which was at first very surprising.  While searching
for its physical origin, we found that it was due to changes in
the size distribution of clusters of lattice spins as the
transition temperature is approached ; it is known indeed from the
theory of critical phenomena that
clusters of up (resp. down) spins of every size form when $T$
approaches $T_c$ in a critical phase transition. At $T_c$, the
distribution of cluster sizes displays clusters of various sizes,
as can be seen from Fig.~\ref{fig:clus} (more details on the
cluster construction algorithm will be given below). As a result,
the conductivity is drastically lower than in the ordered phase
since itinerant electrons have to steer around large clusters in
order to go through the entire structure. Indeed thermal
fluctuations are still not large enough to allow the itinerant spin to
overcome the energy barrier created by the opposite orientation of
the clusters; this is all the more influential that we fixed an
infinite spin-flip time, and this forbids the itinerant electron
to reverse its orientation in order to reduce energy barriers.

\item Below $T_c$, there is no effect of magnetic layer thickness
on $R$. However, for $T>T_c$, the larger thickness yields a
smaller $R$. This can be explained by the effect of interfaces at
nonmagnetic spacers: near $T_c$ the lattice spins at those
interfaces are more strongly disordered  than the bulk lattice
spins, they therefore enhance the resistance.  The importance of
this contribution to the enhancement of the total resistance
depends on the ratio of interface spins to bulk spins. This ratio
becomes smaller when the magnetic layer thickness is larger.

\end{itemize}

Far above $T_c$, most clusters have a small size, so that the
resistivity is still quite large with respect to the low-$T$
phase. However, a few facts account for the decrease of the
resistivity as $T$ is increased: (i) thermal fluctuations are now
sufficient to help the itinerant spin overcome energy barriers
that may occur when it bumps into islands of opposite orientation;
(ii) the cluster size is now comparable with the radius $D_1$ of
the interaction sphere, which in turns reduces the height of
potential energy barriers.

We have pitted this interpretation by first creating an artificial
structure of alternate clusters of opposite spins and then injecting
itinerant spins into the structure.  We observed that itinerant
spins do advance indeed more slowly than in the completely disordered
phase (high-$T$ paramagnetic phase). This finding is very
interesting. We believe that it will have other important related physical
effects yet to be discovered.

In order to show the existence of clusters at a given temperature,
we have used the Kopelman algorithm to construct clusters \cite{Hoshen}.  We
show in Fig. \ref{fig:clus} the distribution of cluster size at
different temperatures.  As can be seen, the distribution peak is enlarged with
increasing $T$.

\begin{figure}[ht]
\input{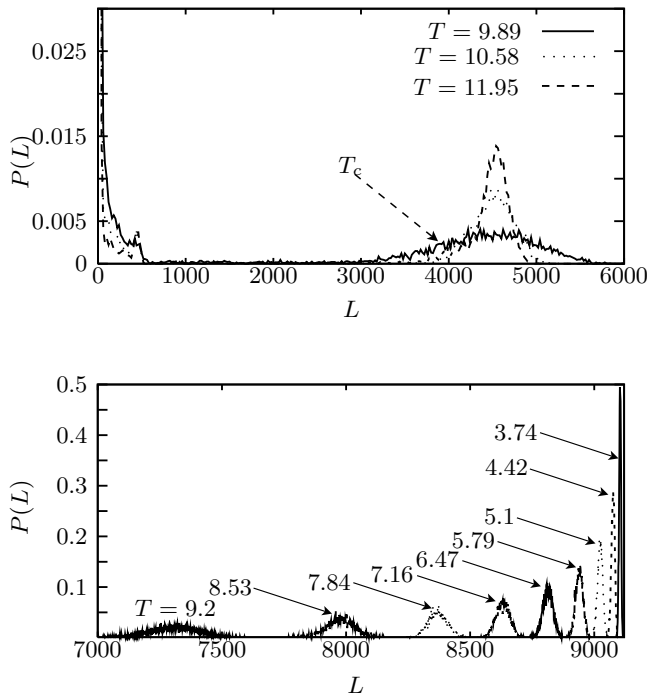} \caption{Distribution $P(L)$ of cluster size $L$
at several temperatures $T$: (a) above $T_c$, (b) below $T_c$.
}\label{fig:clus}
\end{figure}

We plot in Fig. \ref{fig:meansize}(a) the cluster size $A$ as a
function of $T$.  Figure \ref{fig:meansize}(b) shows the $\ln-\ln$
scale of $A$ as a function of $T_c-T$.  The slope is $0.094$
indicating that $A$ does not depend significantly on $T$ at $T<T_c$ as expected.

\begin{figure}
\input{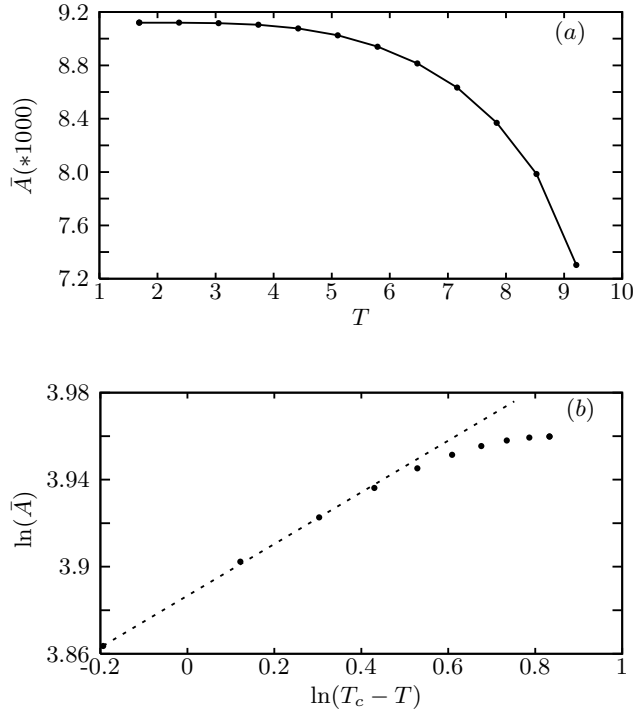}
\caption{a) Average cluster size versus $T$ b) Average cluster
size versus $(T_c-T)$ in the $\ln-\ln$ scale. }\label{fig:meansize}
\end{figure}

At this stage, it is worth to make a comparison with a recent
experiment performed on Ni$_{81}$Fe$_{19}$/Al$_2$O$_3$ multilayer
by Brucas and Hanson.\cite{Brucas} This system is a magnetic
permalloy/insulator multilayer which is very similar to our model:
magnetic layers of thickness $t$ (in the authors' notation) are
separated by insulator of fixed thickness of 16 $\mbox{\AA}$.
Measures of magnetization and resistance have been carried out as
functions of $T$ for two thicknesses $t=16$ and $10$ $\mbox{\AA}$.

For $t=16$ $\mbox{\AA}$, X-ray reflectivity, transmission electron
microscopy and Kerr measurements have shown that the magnetic
layers are ferromagnetic with $T_c\simeq 225$ K. They found that
(see Fig. 2a of Ref. \onlinecite{Brucas}) the resistance is very
small at low $T$ (except for very low $T$), increases slowly with
increasing $T$, makes a jump at $T_c$ and saturated at room
temperature 300 K. This behavior is very similar to what we
obtained in the present paper. We note however that our model
gives a sharper jump than experimental data. This is due to the
perfect crystalline structure (without impurities and defects) of
our model which is certainly not the case of the experimental
system. Besides,  at very low $T$ ($<$ 25 K), due to thermally
frozen dynamics, experimental measures show an anomaly also very
similar to MC results at very low $T$: the decrease of $\lambda$
with decreasing $T$ shown in Fig. \ref{fig:MFP} at T$<$4 means an
increase of $R$ with decreasing $T$. Both experimental and
theoretical systems show therefore a long-relaxation effect due to
finite observation time.

 For $t=10$ $\mbox{\AA}$, the magnetic layers of the
experimental system are in fact composed of superparamagnetic
domains. In contrast to the case of $t=16$ $\mbox{\AA}$, the
resistance in the case of $t=10$ $\mbox{\AA}$ decreases with
increasing $T$ (see Fig. 2b of Ref. \onlinecite{Brucas}).   It is
interesting to note that the experimental system in this case,
which is composed of superparamagnetic domains, is equivalent to
our model in the paramagnetic region above $T_c$ where the
existence of domains of clusters is shown above. The behavior of
the resistance observed in our model for $T>T_c$ is in excellent
agreement with experimental data (see Fig. \ref {resist} at
$T>T_c$).  The effect of domains on the resistance discovered in
our present model is thus verified by this experiment.

\begin{figure}
\input{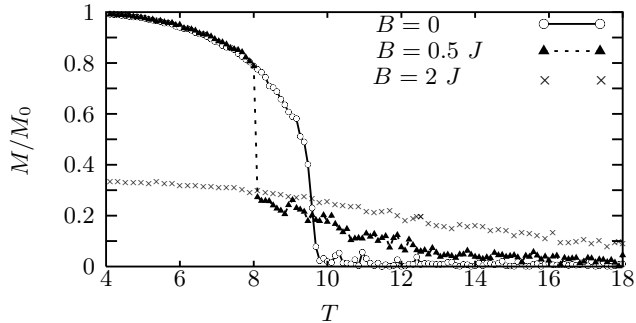}
\caption{Staggered magnetization versus $T$ for several
$B$.}\label{fig:AIM(T)}
\end{figure}

Finally we show the effect of a magnetic field $B$ applied in the
$z$ direction.  If the inter magnetic film coupling is $J_i=1$,
then in the ground state, we need a critical field $B_c=2$ to
align all spins in the $z$ direction.  We show in Fig.
\ref{fig:AIM(T)} the lattice staggered magnetization at $B$=0, 0.5
and 2. As seen, for $B=2$ all lattice spins are aligned in the $z$
direction at low $T$: the staggered magnetization is then equal to $1/3$ at $T=0$.

An applied field much smaller than $B_c$ is expected not to modify
significantly the itinerant spin current at $T\ll T_c$.


%

\begin{figure}
\input{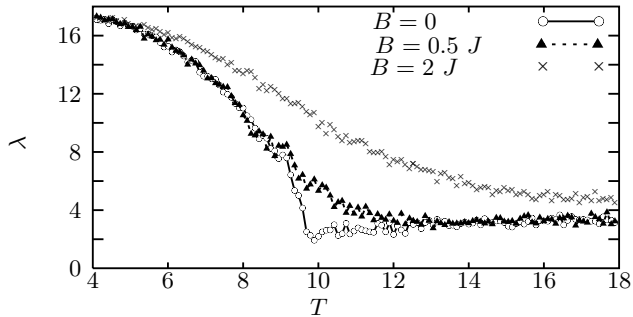}
\caption{Mean free path versus $T$ for several values of $B$.  See text for comments}\label{fig:MF(T)}
\end{figure}

In order to show the effect of the magnetic field strength, we
define the following quantity termed as "magnetoresistance" hereafter
\begin{equation}
Q(B)=\frac{\lambda (B)-\lambda (B=0)}{\lambda (B=0)}\label{eqn:Q}
\end{equation}

\begin{figure}
\input{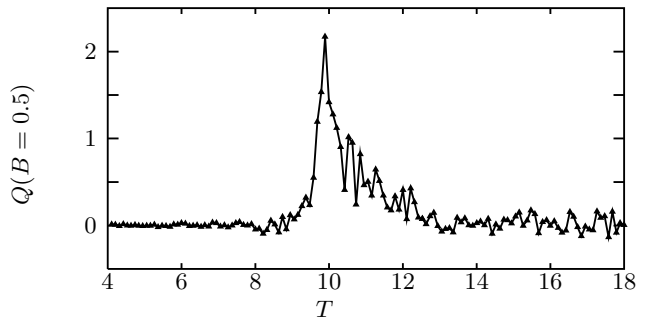}
\caption{Magnetoresistance versus $T$ for a low field $B=0.5$. Note the peak at
$T$ slightly larger than $T_c$.}\label{fig:MRB1}
\end{figure}

 We now show in Fig. \ref{fig:MRB1}
the magnetoresistance in a weak field as a function of $T$. At low $T$,
no significant magnetoresistance is expected since itinerant spins are parallel to
lattice spins. The same is observed at $T$ much larger than $T_c$:
the lattice spins are parallel to the applied field, so itinerant
spins will go through the lattice without resistance.  However, at
$T$ slightly larger than $T_c$ we observe a large peak of the
resistance. This peak has the same origin as that observed in Fig.
\ref{resist}, namely it is due to the existence of the structure
of domains in the transition region.

For large fields, the  same is observed except that the peak is wider and stronger as seen
in Fig. \ref{fig:MRB1} for $B=2$.

\begin{figure}
\input{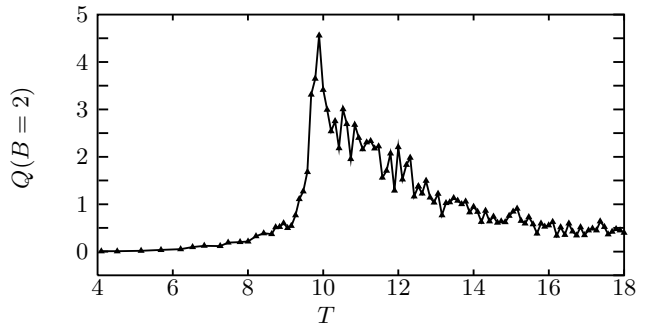}
\caption{Magnetoresistance versus $T$ for a large field $B=2$. Note the peak is larger and higher than
that observed in Fig. \ref{fig:MRB1}.}\label{fig:MRB2}
\end{figure}

\section{Concluding remarks}
We have studied, by means of  MC simulations, the transport of
itinerant spins interacting with localized lattice spins in a
trilayer system of FCC lattice structure in the CIP configuration.
Various interactions have been taken into account.  We found that
the spin current is strongly dependent on the lattice spin
ordering: at low $T$ itinerant spins whose direction is parallel
(antiparallel) to the lattice spins yield a strong (weak) current.
At high temperatures, the
lattice spins are disordered, the current of itinerant spins is
very weak and does not depend on the input orientation of
itinerant spins.  As a consequence, the resistance is very high at
high $T$. We would like to emphasize here a striking effect
found in the transition region between low-$T$ ordered phase
and high-$T$ paramagnetic phase: at $T$ slightly higher
than $T_c$, we discovered a
peak  of the resistance due to the existence of domains of lattice
spins.   Such an existence of domains in the critical temperature region is
well known from the theory of critical phenomena,
 but no one has
expected that it would play an important role in the spin transport.
While writing this paper, we discovered a just-appeared
experimental paper\cite{Brucas} which supports our finding on the
effect of domains in the resistance behavior.

We have also investigated the effects  on the spin current of
different parameters which enter in our model: nonmagnetic spacer
thickness, interaction range, electric field, and magnetic field.
Physical meaning of each of these effects has been discussed. Let
us note that so far, except Ref. \onlinecite{Brucas}, most
magnetoresistance experiments have been
performed as a function of an applied magnetic field, at a given
temperature. While, in our present study, we have considered the
effect of the lattice ordering on the spin current.  We think that
in the light of the results obtained here, more experiments should
be performed to investigate the effect of multilayer ordering on
the spin transport.  As a final remark, we note that the CPP case
is perhaps more difficult to study because effects from non
magnetic spacers as well as  from impurities and roughness at
interfaces will fully set in. Work is now in progress to study
that case.

{}

\end{document}